\documentclass{statsoc}

\usepackage[a4paper]{geometry}
\usepackage{graphicx}
\usepackage[textwidth=8em,textsize=small]{todonotes}
\usepackage{amsmath}
\usepackage{natbib}
\usepackage{etoolbox}
\usepackage{mwe}

\usepackage{eucal}
\usepackage{multicol}
\usepackage{accents}
\usepackage{dutchcal}
\usepackage{bbm}

\newcommand{\ut}[1]{\underaccent{\tilde}{#1}}
\renewcommand{\vec}[1]{\ut{#1}}

\DeclareMathOperator*{\argmin}{arg\,min}

\newcommand{\vechatmu}{\mbox{\large$\vec{\hat{\mu}}$}}

\newcommand{\spe}{\vec{\mathbcal{E}}}
\newcommand{\vece}{\mbox{\Large$\vec{e}$}}
\newcommand{\vecomega}{\mbox{\Large$\vec{\omega}$}}
\newcommand{\vecdelta}{\mbox{\large$\vec{\delta}$}}

\newcommand{\vecw}{\mbox{$\vec{\textbf{w}}$}}
\newcommand{\vecU}{\mbox{\Large$\vec{u}$}}
\newcommand{\vecy}{\mbox{$\vec{y}$}}

\newcommand{\Matmu}{\textbf{\Large$\mu$}}
\newcommand{\Matphi}{\textbf{\Large$\phi$}}

\newcommand{\MatSigma}{\textbf{\large$\Sigma$}}

\newcommand{\Matw}{\textbf{w}}
\newcommand{\MatD}{\textbf{D}}
\newcommand{\MatQ}{\textbf{Q}}
\newcommand{\MatS}{\textbf{S}}

\usepackage{xspace}

\title[Statistical Downscaling with BGL]{Statistical Downscaling of Model Projections with Multivariate Basis Graphical Lasso}

\author[Ekanayaka {\it et al.}]{Ayesha Ekanayaka}
\address{University of Cincinnati,
Cincinnati,
USA.}
\email{ekanaykk@mail.uc.edu}
\author{Emily Kang}
\address{University of Cincinnati,
Cincinnati,
USA.}
\author{Amy Braverman}
\address{Jet Propulsion Laboratory, California Institute of Technology,
Pasadena,
USA.}
\author[Ekanayaka et al.]{Peter Kalmus}
\address{Jet Propulsion Laboratory, California Institute of Technology,
Pasadena,
USA.}

\begin{document}
\begin{abstract}
We describe an improved statistical downscaling method for Earth science applications using multivariate Basis Graphical Lasso (BGL). We demonstrate our method using a case study of sea surface temperature (SST) projections from CMIP6 Earth system models, which has direct applications for studies of multi-decadal projections of coral reef bleaching. We find that the BGL downscaling method is computationally tractable for large data sets, and that mean squared predictive error is roughly $8\%$ lower than the current state-of-the-art interpolation-based statistical downscaling method. Finally, unlike most of the currently available methods, BGL downscaling produces uncertainty estimates. Our novel method can be applied to any model output variable for which corresponding higher-resolution observational data is available.
\end{abstract}

\section{Introduction}

Global climate models (GCMs) produce projections on relatively coarse spatial scales of up to $\sim$100\,km, due to computational limitations, the need for global coverage, and the need for model runs out to at least 2100. However, many applications require projections at significantly higher resolution. This gulf in resolution can be bridged by downscaling, a means for obtaining fine-resolution climate projections from coarse-resolution GCMs. Downscaling comes in two main varieties: dynamical downscaling (DD), in which the coarse-resolution projections are used as inputs to drive regional models that produce fine-resolution results; and statistical downscaling (SD), in which statistical relationships are derived between the coarse-resolution GCM projections and fine-resolution observations. Here, we introduce a novel statistical downscaling method. We developed our method in the application to downscaling sea surface temperatures (SSTs) for use in understanding projected tropical coral reef severe bleaching. 

Coral reefs are critically threatened by rapidly increasing ocean warming\,\citep{Hughes2003, Hoegh-Guldberg2007, Gattuso2015, IPCCSPM2018} in addition to local stresses such as destructive fishing practices and coastal development. Anomalously high sea surface temperatures (SSTs) contributes to severe coral bleaching, in which corals expel their photosynthetic algal partners, and can even cause more immediate thermal death\,\citep{Carilli2009, Hughes2018b}. 

There is increasing interest in use of projections of SST from global climate models and Earth system models (GCMs and ESMs) to infer possible futures of coral reefs\,\citep{Hoegh1999, Hooidonk2013, Hooidonk2016}. However, to be useful either for research or for local conservation management, these projections must capture  spatial variation on scales much finer than current GCM and ESM resolutions of $\sim$100\,km\,\citep{Hooidonk2016}. These coarse-scale projections may be downscaled via regional models (dynamical downscaling) or via high-resolution observational data sets (statistical downscaling). For the coral reef application, statistical downscaling of SST time series to the 4\,km scale compares well to dynamical downscaling, which is still computationally prohibitive\,\citep{Hooidonk2015, Hooidonk2016}.

Statistical downscaling is performed by establishing a strong link between coarse resolution model projections and fine-resolution observations. In Earth science applications, downscaling methods referred to as statistical downscaling do not always utilize statistical models. For example, statistics from both model outputs and observations or reanalysis can be used to build the link, which can be deterministic and not be based on a statistical model; examples of this include Model Output Statistics (MOS) and Perfect Prognosis (PP) \citep{Schmidli,Soares}. Prior downscaling approaches typically do not handle spatial dependency structures; often, smoothing or interpolation methods are used \citep{Timm,AHMED2013320}. For example, a recent work demonstrated a downscaling approach based on bilinear interpolation and produced SSTs at 4\,km\, resolution; however, the method did not account for spatial dependencies, nor did not provide associated uncertainty measures \citep{van2015}. Popular probabilistic approaches which do provide uncertainty estimates include fitting regression-based models \citep{Mishra2014MultipleLR} and generalized linear models relating predictand with potential environmental covariates \citep{beecham2014statistical}. Yet, there are some downscaling strategies which are based on statistical models (usually regression models) but do not provide an uncertainty measure, or only provide uncertainty measures at the observation location (e.g., monitoring stations) \citep{Sachindra,GAITAN20192778}. \citep{berrocal2010spatio} proposed another linear regression model directly relating fine-scale observations to numerical model outputs as a fully model-based solution to the SD problem .The proposed spatio-temporal model in this work assumes spatially and temporally-varying regression coefficients but computationally prefer independence across time. Further, computation involves MCMC simulations which can be challenging for a global scale SD. Another interesting two stage SD method was proposed by \cite{Poggio2015DownscalingAC} using  Generalized Additive Models (GAMs) to first model the trend and then use kriging to model for residuals. Their process is repeated iteratively subjected to a pre-specified stopping criteria and hence can be computationally challenging.

In most Earth science applications for downscaling, significant spatio-temporal dependencies exist, and can be utilized to improve downscaling performance. Here, we propose such a statistical downscaling technique. We compare the method with the standard downscaling method \citep{van2015} and a two-stage downscaling method \citep{Poggio2015DownscalingAC} implemented using local approximate Gaussian Process (laGP) \citep{Robert2015}, via application to SST projections for coral reef studies.

BGL is a modeling framework developed for highly-multivariate processes observed at large number of spatial locations with non-stationary data and inter-variable dependencies \cite{BGL}. With BGL, we propose a computationally efficient statistical downscaling technique accounting for spatial dependence,  providing associated uncertainty, that can be used in Earth science contexts with large datasets. Here, we demonstrate it by downscaling SST projections to the 1\,km scale, a 16-fold resolution improvement over a prior interpolation-based SST downscaling method. We perform a representative case study and validation in the Great Barrier Reef (GBR) region. According to \citep{hessd-6-6535-2009} uncertainties of downscaling results can arise from (1) parent GCM; (2) climate change emission scenarios; (3) observed data; and (4) the method used for downscaling. Therefore, a proper probabilistic assessment of uncertainty is a demanded skill in the context of statistical downscaling. Thus, the advantages of our novel downscaling method include significant skill improvement and uncertainty quantification. 

Section\,\ref{section:data} describes the data and GCM models used in our study. Section\,\ref{section:methodology} describes our downscaling methodology. Section\,\ref{section:results} validates our method using representative results from the GBR case study, quantifies skill improvement relative to the prior state of the art, and introduces the uncertainty quantification from our methodology. Section\,\ref{section:conclusion} provides discussion and conclusion.

\section{Data and model output} \label{section:data}
We use monthly averaged NASA/JPL Multiscale Ultrahigh Resolution (MUR, \cite{MUR}) satellite SST data at 1\,km resolution from June 2002 to December 2020 (a total of 223 months). We use a 4\,km resolution reef mask from the NOAA Coral Reef Watch thermal history product, v1.0\,\citep{Heron2016} to determine the locations of coral reefs in the global ocean. We use monthly SST output from 19 Coupled Model Intercomparison Project Phase 6 (CMIP6) GCM models under the Shared Socioeconomic Pathways (SSPs)  SSP126\,\citep{SSPs}. The model time series are re-gridded to a common $1^\circ$ grid, and run from from June 2002 to December 2099 (1183 months). At each grid location, we take the mean of these model time series.

The study area includes a total of selected 309,700 1\,km MUR pixels and 35 $1^\circ$ coarse grid cells. Only pixels identified as corals and their adjacent neighbours were used for the analysis.

\section{Methodology} \label{section:methodology}

 Let $W_{it}(A_{k})$ be the averaged GCM outputs at coarse grid cell $A_{k} $ and  let $w_{it}(s_{n})$ be the averaged GCM output \textbf{interpolated} to MUR pixel $s_{n}$ where, $n=1,\ldots,309700$ is for all the MUR pixels, $k=1,\ldots,35$ is for all the coarse grid cells in spatial domain $\mathcal{D}$ and $t=1,\ldots,T_{i}$; $T_{i}$ is the total number of months from June 2002 to December 2099 for $i=1,2,\ldots,12$ denoting 12 months. Let $y_{it}(s_{n})$ be the monthly averaged observational SST at MUR location $s_{n}$ where, $t=1,\ldots,T_{o_{i}}$; $T_{o_{i}}$; is the total number of observational months from June 2002 to December 2020 (MUR data available only from 2002 to 2020). Then our downscaling method is performed in two stages, assuming an additive model for fine-resolution observations. i.e. we assume,
 
\begin{equation}
y_{it}(s_{n})=\mu_{it}(s_{n})+\epsilon_{it}(s_{n})
\end{equation}

 In the first stage, we estimate large-scale spatio-temporal variations, often referred to as the trend or mean component ($\mu$), using a deterministic approach. We notice that this mean component is capable of capturing an extensive amount of large-scale variations. However, we hypothesized that significant fine-scale variations remained unexplained, and accounting further for the small-scale variations will help increase the accuracy in downscaled SSTs. Therefore in the second stage, we propose a further step to model for the remaining small-scale variations.

\subsection{Stage 1: Estimating trend component}

The deterministic procedure for trend estimation consist of steps. The first step is to subtract model climatology \{$\bar{W_{i}}(A_{k})$)\} which is defined as,
\begin{equation}
\bar{W_{i}}(A_{k})=\frac{\sum_{t=1}^{T_{o_{i}}}W_{it}(A_{k})}{T_{o_{i}}}
\end{equation} from model data. Then the resulting time series are interpolated from model pixels to MUR pixels using bivariate interpolation. Finally, the trend is estimated by adding interpolated values to the MUR climatology \{$\bar{y_{i}}(s_{n})$\} which is defined as,

\begin{equation}
\bar{y_{i}}(s_{n})=\frac{\sum_{t=1}^{T_{o_{i}}}y_{it}(s_{n})}{T_{o_{i}} }
\end{equation}



\subsection{Stage 2: Model for small-scale variations}

In stage 2, we propose a further step to capture remaining small-scale variations using BGL. We begin assuming the vector, $\vece(s_{n})=\Big(e_{1}(s_{n}),e_{2}(s_{n})\Big)^T$ where $e_{1}(s_{n})=w_{it}(s_{n})-\frac{\sum_{t=1}^{T_{o_{i}}}w_{it}(s_{n})}{T_{o_{i}}}$ and $e_{2}(s_{n})=y_{it}(s_{n})-\hat{\mu}_{it}(s_{n})$ for fixed $i,t$ and any $s_n\in \mathcal{D}$ follow a bivariate Gaussian process. We further assume that this vector $\vece(s_{n})$, can be additively modelled using a spatially correlated stochastic process $\vecU(s_{n})$ and a white noise process $\vecdelta(s_{n})$. i.e.
\begin{equation}
\vece(s_{n})=\vecU(s_{n})+\vecdelta(s_{n})
\end{equation}

where $\vecU(s_{n})=\Big(u_{1}(s_{n}),u_{2}(s_{n})\Big)^T$ and $\vecdelta(s_{n})=\Big(\delta_{1}(s_{n}),\delta_{2}(s_{n})\Big)^T$ is a mean zero white noise process with  $Cov(\delta_{1}(s_{n}),\delta_{2}(s_{n}))=diag(\tau_{1}^2,\tau_{2}^2)$. Then the BGL idea in \cite{BGL} relies on  basis expansion of the components of spatially correlated stochastic process. We use empirical orthogonal functions (EOFs) as the basis functions. In the current study, we only have limited number of observational months. Therefore we combine months into seasons and fit four different BGL models for each season. The total number of available EOFs is, $T=T_{o_{1}}+T_{o_{2}}+T_{o_{3}}$. Then the spatially correlated stochastic process $u_{j}(s_{n})$ is further expressed as,

\begin{equation}
u_{j}(s_{n})=\sum_{l=1}^{L}\omega_{jl}\phi_{l}(s_{n})+\sum_{l=1}^{T-L}\nu_{jl}\psi_{l}(s_{n})
\end{equation}
where $\phi_{l}(s_{n})$ and $\psi_{l}(s_{n})$ are  basis functions and $\omega_{jl}$ and $\nu_{jl}$ are the respective coefficients.
Here we assume that $\sum_{l=1}^{L}\omega_{jl}\phi_{l}(s_{n})$ is a stochastic term and the $\sum_{l=1}^{T-L}\nu_{jl}\psi_{l}(s_{n})$ term is deterministic. We follow the methodology presented in \cite{MISR} to separate basis functions into deterministic and stochastic terms. We use ordinary least squares estimates for deterministic coefficients and then for the stochastic coefficients we further assume,
\begin{equation}\label{Omega_l}
\vecomega_{l}=\Big(\omega_{1l},\omega_{2l}\Big)^T\sim N\Big(0,\MatQ_{l}^{-1}\Big)
\end{equation}
and wish to obtain a sparse non-parametric estimate from BGL for the precision matrix $\MatQ$ of vector of all the stochastic coefficients $\vecomega=(\vecomega_{1},..,\vecomega_{L})^T$, assuming $\MatQ=diag(\MatQ_{1},..,\MatQ_{L})$. This is achieved by first defining a new vector $\spe=(\vece(s_{1})^T,..,\vece(s_{n})^T)^T$ for fixed $i,t$ which lists all the $\vece(s_{n})$ vectors over the spatial domain $\mathcal{D}$. Here, let $\MatSigma=Var(\spe)$
and then assuming each $\spe_{it}$ is a different realization of $\spe$ we see that the joint negative log-likelihood is proportional to,
\begin{equation}
log ( det( \MatSigma))
+\frac{\sum_{i=1}^{3}\sum_{t=1}^{T_{o_{i}}}\spe_{it}^{T}\MatSigma^{-1}\spe_{it}}{T_{o_{1}}+T_{o_{2}}+T_{o_{3}}}
=log ( det( \MatSigma))
+tr(\MatS\MatSigma^{-1})
\end{equation}
where $\MatS=\frac{\sum_{i=1}^{3}\sum_{t=1}^{T_{o_{i}}}\spe_{it}^{T}\spe_{it}}{T_{o_{1}}+T_{o_{2}}+T_{o_{3}}}$. Note that here, $\MatSigma=\Matphi \MatQ^{-1}\Matphi^T +\MatD$, where $\Matphi$ is the basis matrix and $\MatD=diag(\tau_{1}^2,\tau_{2}^2)\otimes I_{n}$.

Then BGL solves $l_{1}-$penalized maximum likelihood equation,
\begin{equation}\label{Likelihood}
\hat{\MatQ} \in \argmin_{\MatQ\succeq 0} log(det(\Matphi \MatQ^{-1}\Matphi^T +\MatD))+tr(\MatS(\Matphi \MatQ^{-1}\Matphi^T +\MatD)^{-1}) +P(\MatQ)
\end{equation} 
where,\vspace{-0.9\baselineskip}
\begin{equation}
P(\MatQ)=P(\MatQ_{1},..,\MatQ_{L})=\lambda\sum_{l=1}^{L}\sum_{i\neq j}|(\MatQ_{l})_{ij}|+\rho\sum_{l=1}^{L-1}\sum_{i\neq j}|(\MatQ_{l})_{ij}-(\MatQ_{l+1})_{ij}|\nonumber
\end{equation}
to obtain the an estimate for the precision matrix $\MatQ$.Here, $\lambda$ is the sparsity inducing penalty and $\rho$ is a fusion penalty which penalize $\MatQ_{l}$ matrices at adjacent levels if the off-diagonals are not similar. Notice that, evaluation of this likelihood equation requires an expensive Choleskey decomposition ($\mathcal{O}(p^3n^3)$).

Using the Sherman-Morrison-Woodbury formula, likelihhod expression in expression \ref{Likelihood} can be re-written reducing likelihood evaluation to $\mathcal{O}(p^3L^3)$ as,
\begin{eqnarray} \label{Likelihood1}
{}&log(det(\MatQ+\Matphi^T\MatD^{-1}\Matphi))-log(det(\MatQ))- \nonumber\\&\  tr(\Matphi^T\MatD^{-1}\MatS\MatD^{-1}\Matphi(\MatQ+\Matphi^T\MatD^{-1}\Matphi)^{-1})+P(\MatQ)
\end{eqnarray}
The block-diagonal structure of $\MatQ$ further reduces matrix computation of size $pL \times pL$ to $L$ computations of $p \times p$ matrices. However, the likelihood expression in expression \ref{Likelihood1} is yet non-smooth and non-convex with respect to $\MatQ$. Thus, authors use difference-of-convex (DC) algorithm where next guess of $\MatQ$ is obtained by solving a convex optimization problem with the concave part linearized at the previous guess $\MatQ^{(j)}$ \citep{BGL}.

Now recall that $e_{1}(s_{n})=w_{it}(s_{n})-\frac{\sum_{i=1}^{3}\sum_{t=1}^{T_{o_{i}}}w_{it}(s_{n})}{T_{o_{1}}+T_{o_{2}}+T_{o_{3}}}$ are available for $t=1,\ldots,T_{i}(>T_{o_{i}})$ but $e_{2}(s_{n})=Y_{it}(s_{n})-\hat{\mu}_{it}(s_{n})$ are only available for $t=1,\ldots,T_{o_{i}}$. With a simple re-arrangement of vector of residuals, we can re-write vector $\spe$ as follows.

\[ 
\spe=\begin{bmatrix}
\spe_{\mathbbm{1}}\\  
\spe_{\mathbbm{2}}\\
\end{bmatrix}
=
\begin{bmatrix}
\Matphi \vecomega_{\mathbbm{1}}\\  
\Matphi \vecomega_{\mathbbm{2}}\\   
\end{bmatrix}
+
\begin{bmatrix}
\vecdelta_{\mathbbm{1}}\\  
\vecdelta_{\mathbbm{2}}\\   
\end{bmatrix}  
=
\begin{bmatrix}
\Matphi A_{1} \vecomega\\  
\Matphi A_{2} \vecomega\\   
\end{bmatrix}
+
\begin{bmatrix}
\vecdelta_{\mathbbm{1}}\\  
\vecdelta_{\mathbbm{2}}\\   
\end{bmatrix}  
\] 
where $\spe_{\mathbbm{1}}=\Big(e_{1}(s_{1}),\ldots,e_{1}(s_{n})\Big)^T$,$\spe_{\mathbbm{2}}=\Big(e_{2}(s_{1}),\ldots,e_{2}(s_{n})\Big)^T$ $\vecomega_{\mathbbm{1}}=A_{1}\vecomega=\Big(\omega_{11},\ldots,\omega_{1L}\Big)^T$,
$\vecomega_{\mathbbm{2}}=A_{2}\vecomega=\Big(\omega_{21},\ldots,\omega_{2L}\Big)^T$
and $\vecdelta_{\mathbbm{1}}=\Big(\delta_{1}(s_{1}),\ldots,\delta_{1}(s_{n})\Big)^T$, $\vecdelta_{\mathbbm{2}}=\Big(\delta_{2}(s_{1}),\dots,\delta_{2}(s_{n})\Big)^T$.

Then, for a future month $t>T_{o_{i}}$ our goal is to obtain $Exp\big[\vecomega_{\mathbbm{2}}|\spe_{\mathbbm{1}}]$ as predicted residuals. Given $\spe_{\mathbbm{1}}$ we first obtain generalized least squares estimates for $\vecomega_{\mathbbm{1}}$.
\begin{equation}
\hat{\vecomega}_{\mathbbm{1}_{GLS}}=\Big(\Matphi \MatSigma_{\mathbbm{1}}^{-1} \Matphi ^{T}\Big)^{-1}\Matphi \MatSigma_{\mathbbm{1}}^{-1}\spe_{\mathbbm{1}}
\end{equation}
where $\MatSigma_{\mathbbm{2}}=Var(\spe_{\mathbbm{1}})=\Matphi A_{1}\MatQ^{-1}A_{1}^{T}\Matphi^{T}+\MatD_{\mathbbm{1}}$ and $\MatD_{\mathbbm{1}}=\tau_{1}^{2}I_{n}$. Recall from equation \ref{Omega_l} we assume a bivariate Normal distribution for the vector of stochastic coefficients $\Big(\omega_{1l},\omega_{2l}\Big)^T$ at each level $l$. Thus, using the law of total expectation we can obtain predicted vector of stochastic coefficients $\hat{\vecomega}_{\mathbbm{2}}$ as,
\begin{equation}
    \hat{\vecomega}_{\mathbbm{2}}=Exp\Big[ \vecomega_{\mathbbm{2}}|\spe_{\mathbbm{1}}\Big]=Exp\Big[Exp\Big[\vecomega_{\mathbbm{2}}|\spe_{\mathbbm{1}},\vecomega_{\mathbbm{1}}\Big]|\spe_{\mathbbm{1}}\Big]
\end{equation}

with conditional variance,
\begin{equation}
    Var\Big[ \vecomega_{\mathbbm{2}}|\spe_{\mathbbm{1}}\Big]=Exp\Big[Var\Big[\vecomega_{\mathbbm{2}}|\spe_{\mathbbm{1}},\vecomega_{\mathbbm{1}}\Big]|\spe_{\mathbbm{1}}\Big]+Var\Big[Exp\Big[\vecomega_{\mathbbm{2}}|\spe_{\mathbbm{1}},\vecomega_{\mathbbm{1}}\Big]|\spe_{\mathbbm{1}}\Big]
\end{equation}
Finally, we obtain the vector of downscaled SSTs for a future month $t>T_{o_{i}}$ as,
\vspace{-0.1\baselineskip}
\begin{equation}
\hat{\vecy}=\vechatmu+\hat{\spe}_{\mathbbm{2}}\nonumber
\end{equation}

where, $\hat{\vecy}=\Big(\hat{y}(s_{1}),\ldots,\hat{y}(s_{n})\Big)^T$, $\vechatmu=\Big(\hat{\mu}(s_{1}),\ldots,\hat{\mu}(s_{n})\Big)^T$ and $\hat{\spe}_{\mathbbm{2}}=\Matphi\hat{\vecomega}_{\mathbbm{2}}$.

\section{Results} \label{section:results}

We performed validation leaving out the three years from 2018 to 2020 to assess performance of the proposed method.  We calculated averaged Mean Squared Error(MSE) to assess prediction accuracy over time as well as over the space. We also used Structural Similarity Index Measure (SSIM) to measure similarity between observational MUR SST maps and the downscaled SST maps \citep{SSIM}. We compare performances of our method with two previous SD methods; the interpolation based standard statistical downscaling method used by e.g., \citep{van2015}, and the two-stage method of \citep{Poggio2015DownscalingAC} but replacing the GAM estimated trend with our trend from Stage 1 and replacing kriging step with local approximate Gaussian Process (laGP) by \citep{Robert2016}. The three input variables used were longitude, latitude and the difference between interpolated GCMs and the trend. i.e $\Matw-\hat{\Matmu}$ where $\Matw=\Big(\vecw_{11},\ldots,\vecw_{1T_{o_{1}}},\vecw_{21},\ldots,\vecw_{2T_{o_{2}}},\vecw_{31},\ldots,\vecw_{3T_{o_{3}}}\Big)$, $\vecw_{it}=\Big(w_{it}(s_{1}),\ldots,w_{it}(s_{n})\Big)^T$.

\subsection{Validation with MUR data}

\begin{figure}
\includegraphics[width=0.3\linewidth]{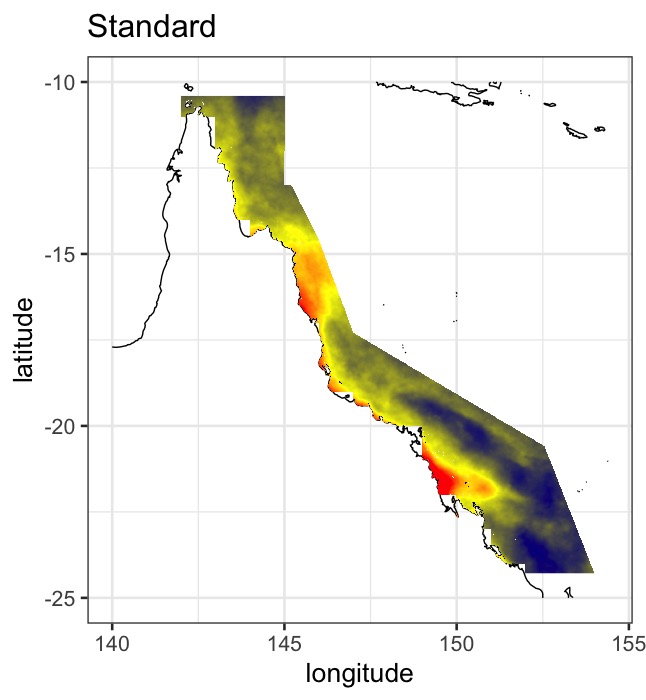}
\includegraphics[width=0.3\linewidth]{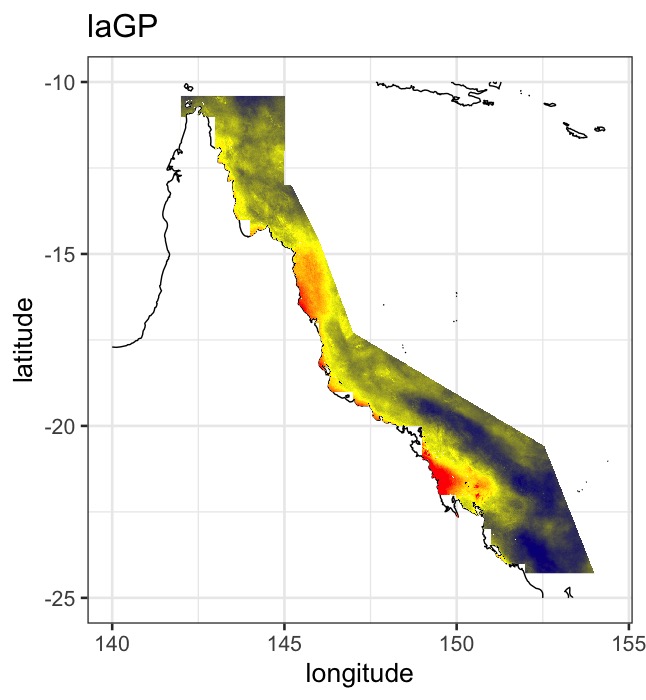}
\includegraphics[width=0.3\linewidth]{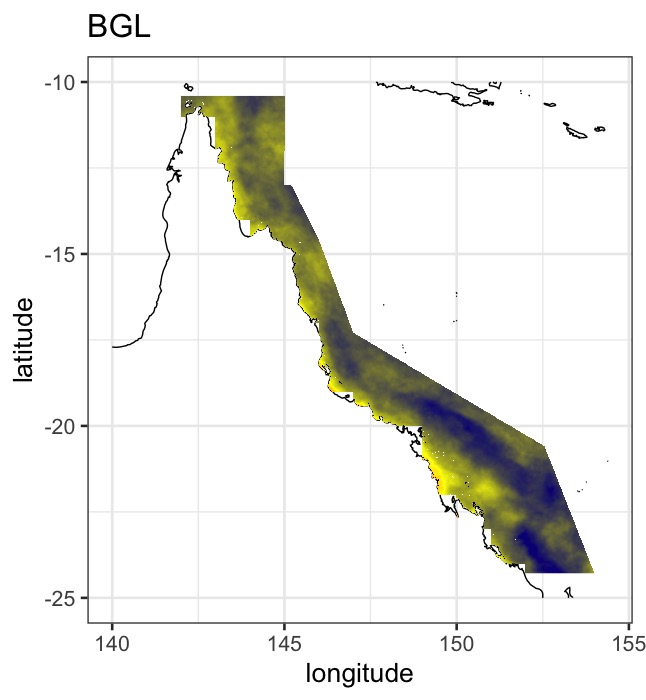}
\\
\centering
\includegraphics[width=0.45\linewidth]{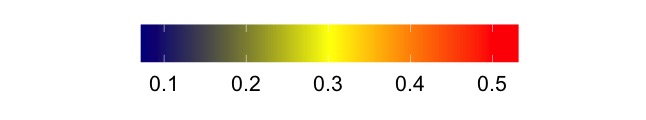}
\caption{\label{MSE_maps.126}}Comparison of MSE maps under ssp126. Notice the improvement in BGL method along the coastline.
\end{figure}

\begin{figure}[ht]
\includegraphics[width=0.45\linewidth]{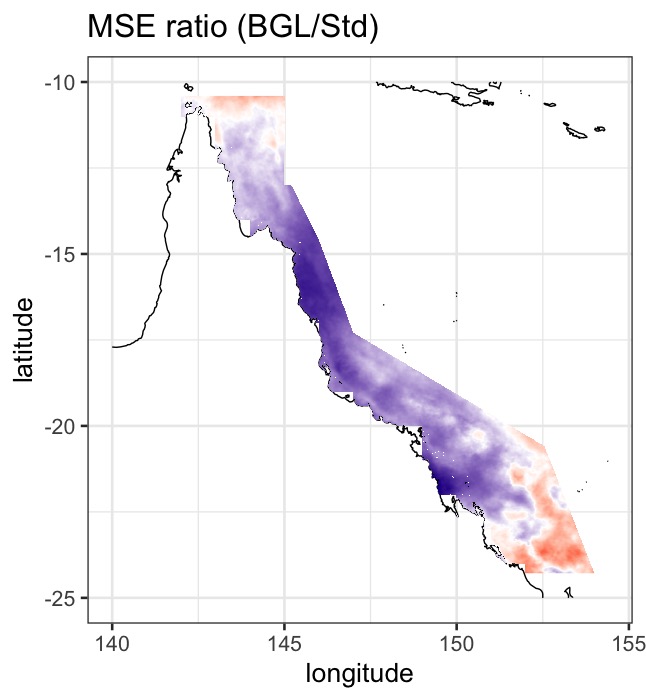}
\includegraphics[width=0.45\linewidth]{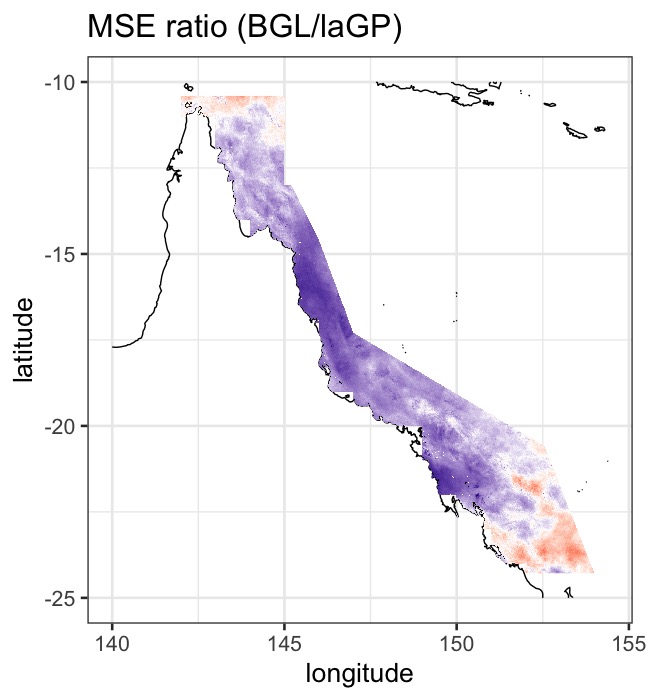}
\\
\centering
\includegraphics[width=0.45\linewidth]{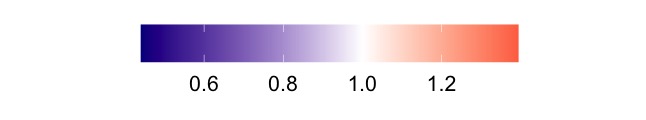}
\caption{\label{MSE.ratio_maps.126}}MSE ratio maps. In general, BGL has the lowest MSE across the region.
\end{figure}

\begin{figure}
\centering
\includegraphics[width=0.35\linewidth]{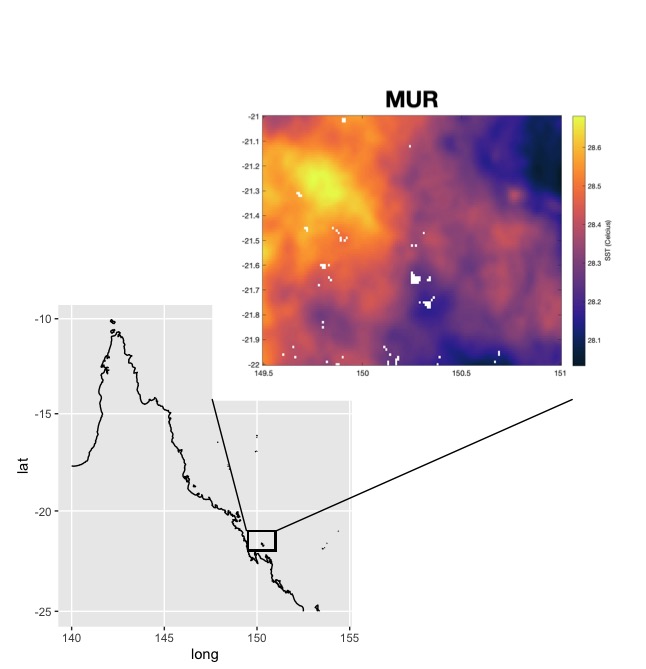}
\\
\vspace{1\baselineskip}
\includegraphics[width=0.22\linewidth]{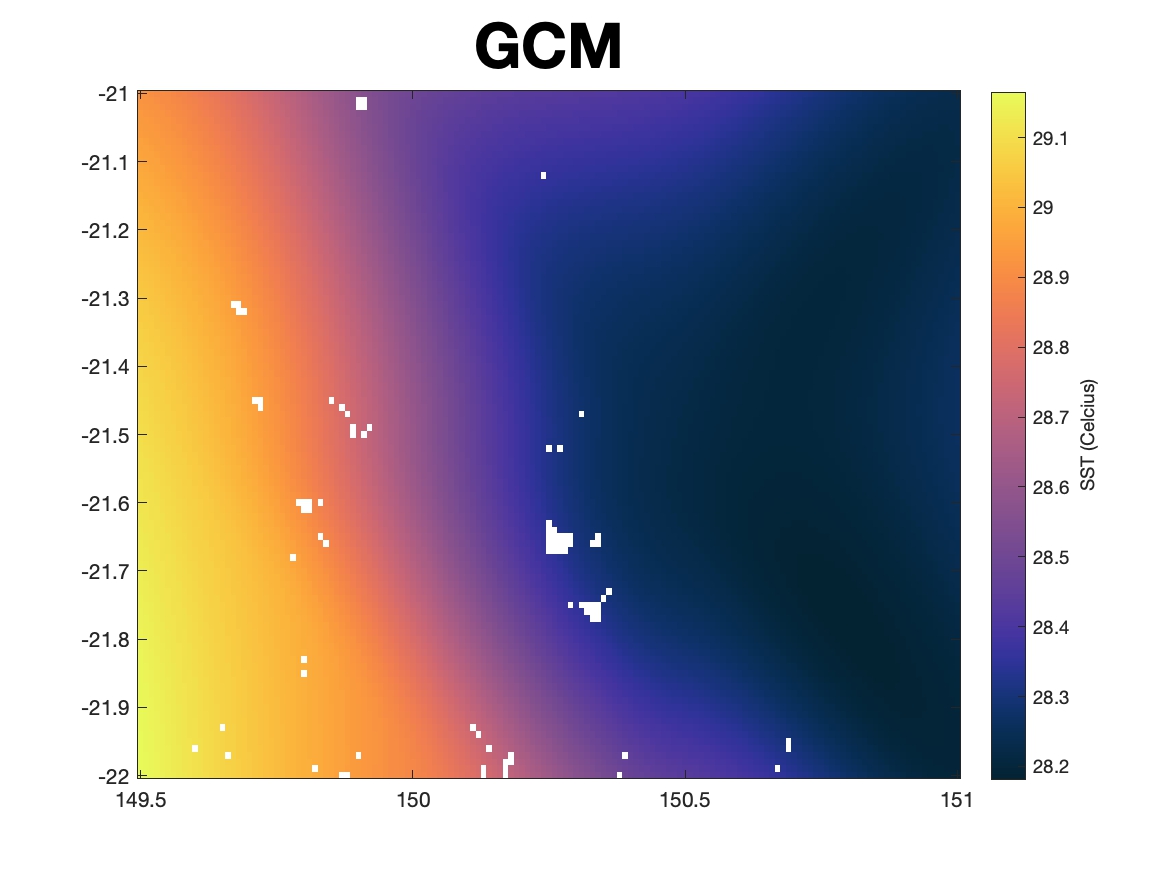}
\includegraphics[width=0.22\linewidth]{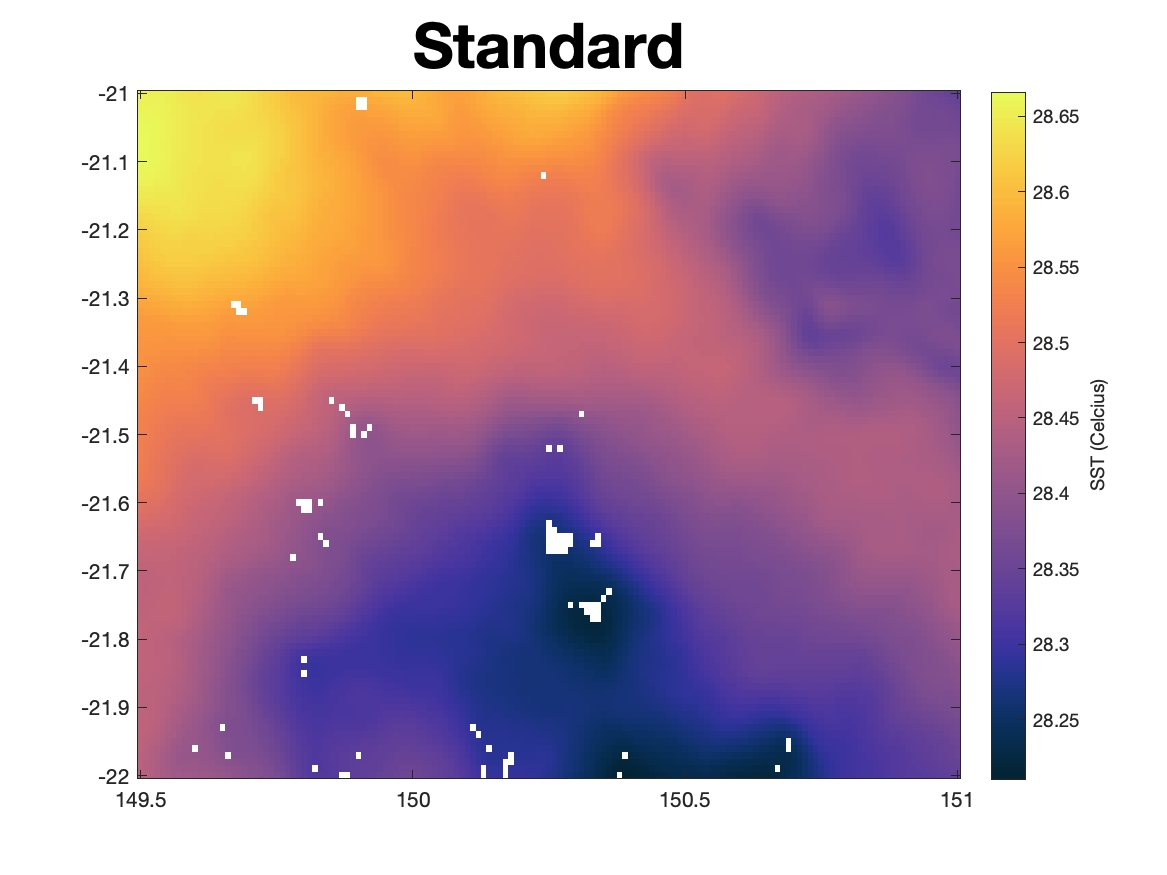}
\includegraphics[width=0.22\linewidth]{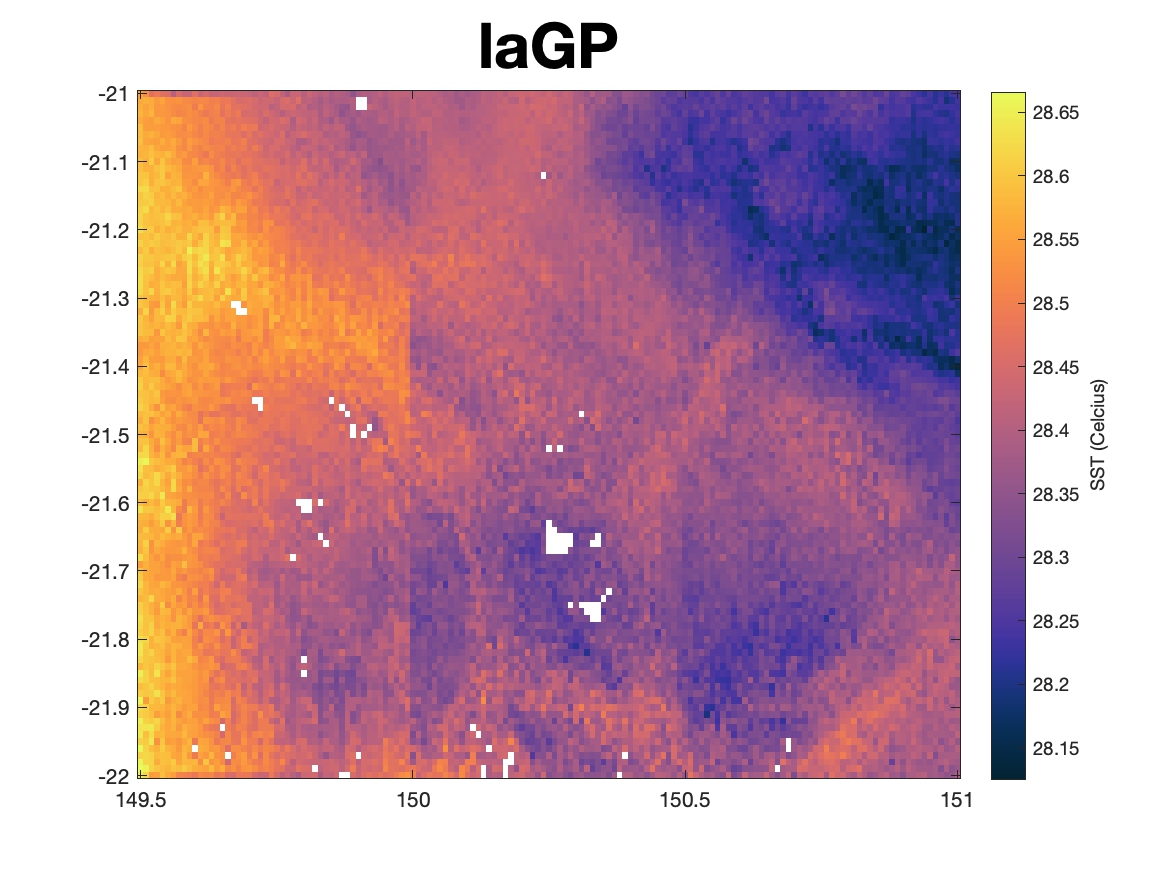}
\includegraphics[width=0.22\linewidth]{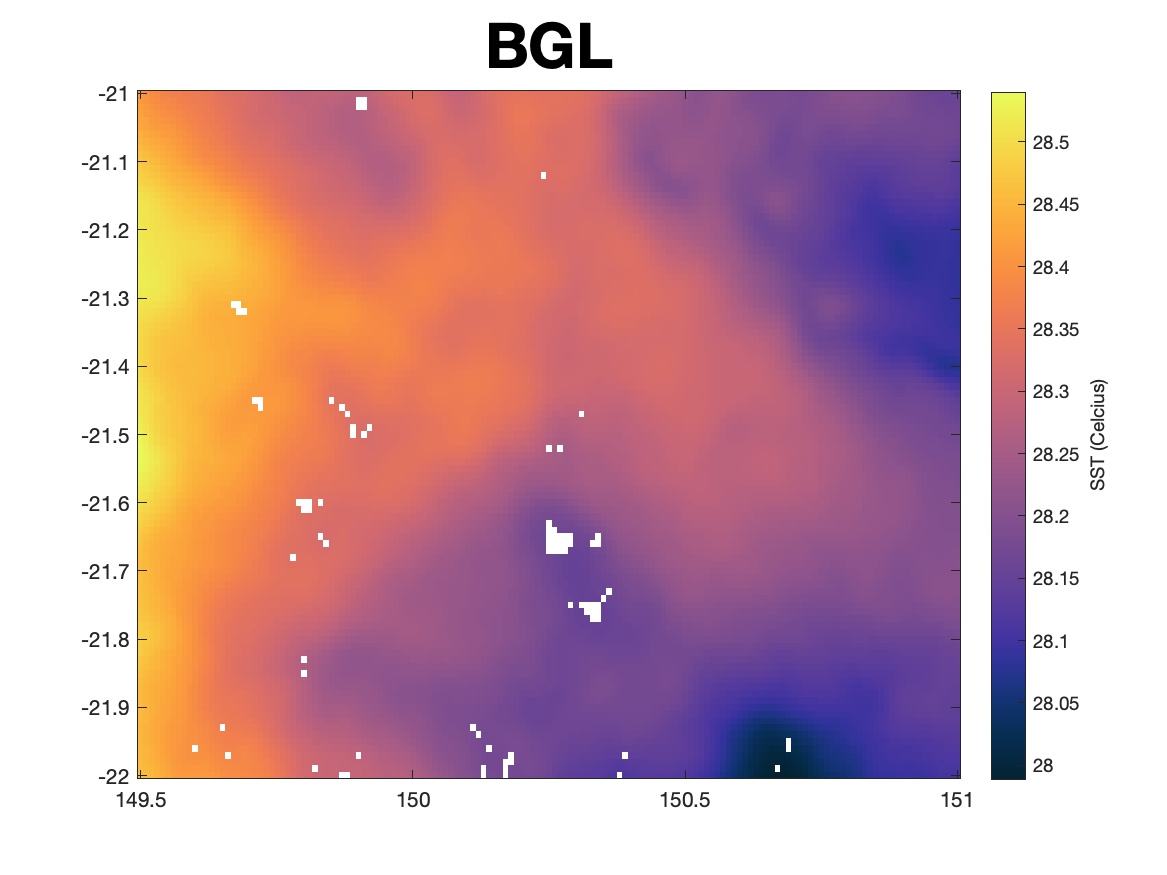}
\caption{\label{SSIM_maps}}SST maps for January 2020 zoomed into a regular grid in the study region. Compare the structural similarity between downscaled SST maps with the observational MUR SST map.
\end{figure}

We left out the years 2018 to 2020 and performed downscaling using MUR data from June 2002 to December 2017. Table \ref{MSE_SSIM_comparison_tab} compares averaged MSE values and averaged SSIM values. From the table we see that overall, BGL method has the lowest MSE and the percentage reduction is significant when it is compared to the standard downscaling method and the laGP method. From Figure \ref{MSE_maps.126}  notice that there is a drastic improvement in MSE specially along the coastline. Figure \ref{MSE.ratio_maps.126} further confirms that BGL has reduced MSE across the region. We list averaged SSIM values calculated zooming into the regular grid shown in Figure \ref{SSIM_maps}. As our study region is in irregular pattern often with empty background, we zoomed into a regular grid to calculate SSIM in a meaning full manner. We use SSIM to measure the structural similarity between observational MUR SST maps and the downscaled SST maps. A number close to 1 indicates a greater similarity. From the table notice that our BGL method has highest averaged SSIM values. 
Notice from the Figure that laGP introduces an usual instability to the SST process, possibly due to its local structure.

\begin{table}
\caption{\label{MSE_SSIM_comparison_tab} MSE and SSIM averaged over years from 2018 to 2020 separated by seasons.}
\centering
\begin{tabular}{|l|l|l|l|l|}
\hline
\multicolumn{5}{|c|}{MSE} \\
\hline
Season & GCM & Standard & laGP  & BGL\\
\hline
Summer & 0.356 & 0.310 & 0.317 & \textcolor{red}{0.277}\\
Autumn & 0.472 & 0.105 & 0.108 & \textcolor{red}{0.093}\\
Winter & 0.396 & 0.305 & 0.307 &  \textcolor{red}{0.288}\\
Spring & 0.380 & 0.212 & 0.221 & \textcolor{red}{0.200}\\
\hline
\textbf{Overall} & 0.401 & 0.233 & 0.238 & \textcolor{red}{0.214}\\
\hline

\end{tabular}
 \begin{tabular}{|l|l|l|l|l|}
\hline
\multicolumn{5}{ |c| }{SSIM} \\
\hline
Season & GCM & Standard & laGP  & BGL\\
\hline
Summer & 0.680 & 0.765 & 0.739 & \textcolor{red}{0.747}\\
Autumn & 0.551 & 0.842 & 0.825 &  \textcolor{red}{0.872}\\
Winter & 0.626 & 0.888 & 0.905 & \textcolor{red}{0.943}\\
Spring & 0.646 & 0.767 & 0.743 & \textcolor{red}{0.875}\\
\hline
\textbf{Overall} & 0.626 & 0.815 & 0.803 & \textcolor{red}{0.859}\\
 \hline
\end{tabular}
\end{table}

\clearpage

\subsection{Future projections}

\begin{figure}[ht]
\includegraphics[width=1\linewidth]{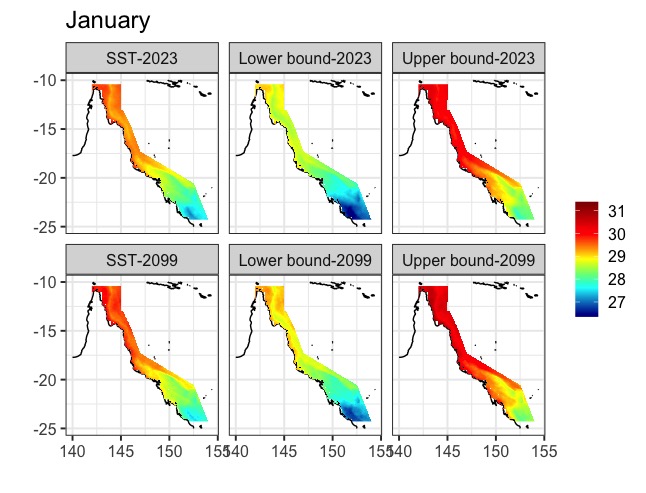}
\caption{\label{Jan2023_2099.126}}Downscaled SSTs with uncertainty boundaries for January 2023 and January 2099.
\end{figure}

In Figure \ref{Jan2023_2099.126} we present a comparison of downscaled SSTs for January 2023 with January 2099 along with the estimated uncertainty from our proposed BGL method. In standard downscaling, the usual practice to obtain an uncertainty measure is to estimate the standard error using MSE which is constant over time. In contrast, our proposed method provides a probabilistic uncertainty which is different for each month. 

\clearpage

\section{Discussion and conclusion} \label{section:conclusion}

We have presented a novel statistical downscaling method that uses BGL for residual estimation. We have demonstrated our BGL downscaling method in a case study of coral reefs under warming SSTs in the Great Barrier Reef region. The novel BGL downscaling method is computationally tractable for large data sets, provides meaningful uncertainty estimates, and reduced overall MSE significantly in this case study. Therefore, it is suitable for a wide range of applications in Earth science and other fields, e.g. for accomplishing the statistical downscaling of coarse-scale global climate model projections using fine-scale observational data.

A hybrid dynamic-statistical downscaling framework also could also be developed that includes global climate model output, regional climate model output, and observations in the downscaling \citep{Walton}. However, this would require a more complicated statistical model to jointly model the three data resources and could be the subject of future work. A possible extension is to generalize the current model to the framework of autoregressive co-kriging for multi-fidelity model output and then consider the observations, regional climate model output, and global climate model output as the high-, medium-, and low-fidelity data, respectively.

\section{Acknowledgments}
Research was carried out at the Jet Propulsion Laboratory, California Institute of Technology, under a contract with the National Aeronautics and Space Administration (80NM0018D0004). Financial and in-kind support for this project was provided by NASA ROSES Sustaining Living Systems in a Time of Climate Variability and Change program, grant number 281945.02.03.09.34; and the University of Cincinnati. We acknowledge the World Climate Research Program’s Working Group on Coupled Modelling, which is responsible for CMIP, and we thank the climate modeling groups for producing and making available their model output. The contents in this manuscript are solely the opinions of the authors and do not constitute a statement of policy, decision or position on behalf of NASA, the Jet Propulsion Laboratory, or the US Government. The authors thank Alex Goodman for developing the Big Climate Data Project (BCDP), and Robert Gramacy for helpful discussion.

\bibliographystyle{apalike}
\bibliography{ref}
\end{document}